\documentclass[12pt]{iopart}
\pdfoutput=1 
\usepackage[utf8]{inputenc}
\usepackage[T1]{fontenc}
\usepackage{epsfig,graphicx,latexsym}

\expandafter\let\csname equation*\endcsname=\relax
\expandafter\let\csname endequation*\endcsname=\relax

\usepackage{amssymb}
\usepackage{amsmath}

\usepackage{xcolor}
\definecolor{darkblue}{rgb}{0,0,0.6}
\definecolor{darkred}{rgb}{0.6,0,0}

\usepackage[colorlinks=true,urlcolor=darkblue,citecolor=darkblue,linkcolor=darkred,hyperfootnotes=false]{hyperref}

\newcommand{\eeta}{{\boldsymbol{\eta}}}

\newcommand{\EE}{\boldsymbol{E}}
\newcommand{\ff}{\boldsymbol{f}}
\newcommand{\FF}{\boldsymbol{F}}
\newcommand{\jj}{\boldsymbol{j}}
\newcommand{\JJ}{\boldsymbol{J}}
\newcommand{\kk}{\boldsymbol{k}}

\newcommand{\uu}{\boldsymbol{u}}
\newcommand{\vv}{\boldsymbol{v}}
\newcommand{\xx}{\boldsymbol{x}}

\graphicspath{{figures/}}

\newcommand{\dd}{\mathrm{d}}

\newcommand{\transp}{^\mathrm{T}}

\renewcommand{\overline}{\bar}

\begin{document}

\title{The conductivity of strong electrolytes from stochastic density functional theory}

\author{V. D\'emery}
\address{PCT, UMR Gulliver 7083, ESPCI and CNRS, 10 rue Vauquelin, 75005 Paris, France}
\ead{vincent.demery@espci.fr}
\author{D. S. Dean}
\address{Laboratoire Ondes et
Mati\`ere d'Aquitaine (LOMA), CNRS, UMR 5798 / Universit\'e de  Bordeaux, F-33400 Talence, France}
\ead{david.dean@u-bordeaux1.fr}
\begin{abstract}
Stochastic density functional theory is applied to analyze the conductivity of strong two species electrolytes at arbitrary field strengths. 
The corresponding stochastic equations for the  density of the electrolyte species are solved by linearizing them about the mean density of ionic species, yielding an effective Gaussian theory. The non-equilibrium density-density correlation functions are computed and the conductivity of the electrolyte is deduced. In the bulk, our results give a simple derivation of the results of Onsager and coworkers who used very different methods. 
The method developed here can also be used to study electrolytes confined in one and two dimensions and interacting via either the three dimensional Coulomb interaction or the Coulomb interaction corresponding to that dimension of space.
\end{abstract}

% Uncomment for keywords
\vspace{2pc}

%\keywords{Correlation functions, Transport processes / heat transfer (Theory), Ionic liquids, electrolyte solutions, liquid metals and charged fluids (Theory)}

%\pacs{}
\maketitle

\section{Introduction}
Onsager's study of the conductivity of strong electrolytes is famous as one of his first major scientific contributions. In this work he showed how the theory developed by Debye and H\"uckel \cite{deb1923} should be modified to take into account the effect of Brownian motion on the ions \cite{ons1927}. For an electroneutral system with two ionic species $\alpha\in\{+,-\}$ with ionic charges $q z_\alpha$ ($z_\alpha$ denoting the signed valency), mobilities $\kappa_\alpha$ and average densities $\overline{\rho}_\alpha$, the naive formula for the  conductivity, which neglects the interaction between the ions, is %given by 
\begin{equation}\label{condbare}
\sigma_0 = q^2\sum_\alpha z_\alpha^2\kappa_\alpha\bar\rho_\alpha.
%q^2 z^2_+ \kappa_+ \overline{\rho}_+  + q^2 z^2_- \kappa_- \overline{\rho}_-.
\end{equation}  

Debye and H\"uckel \cite{deb1923} computed a correction to this bare conductivity due to the flow induced on the solvent by the charge distribution about an ion. Onsager later pointed out how this result is changed (essentially additively at the level of approximation employed) by the Brownian motion of the ions and their mutual interaction \cite{ons1927}. 
Indeed it is clear that even in the absence of the solvent the conductivity will be modified by correlations between the ions: an ion moves in an electric field which is the sum of the uniform applied field and the field generated by the other ions in the system. Onsager then went on to study the conductivity of strong and weak electrolytes in the presence of a finite applied field \cite{hem1996}. 
For strong electrolytes the modification due to the finite electric field is that the Debye screening of charges is killed off by the field and thus the conductivity increases -- the so-called Wien effect \cite{ons1957}.
For weak electrolytes, which consists of free anions and cations  along with bound Bjerrum pairs, the external field has the effect of shifting the chemical equilibrium between ions and Bjerrum pairs (by pulling the pairs apart)
\cite{ons1934}. This shift increases the density of charge carriers,  leading to an  increase in conductivity. The validity of Onsager's calculation for weak electrolytes was verified recently, for a solvent free system, by numerical simulations of 
the second Wien effect \cite{kai2013}.  
Interestingly, Onsager returned regularly to this problem throughout his career, he compared it carefully with experimental results and also developed new mathematics to facilitate the study of these systems \cite{hem1996}. 

In this paper we revisit the problem of the conductivity of strong electrolytes for Brownian systems (we do not consider the effect of a solvent here) by analysing the stochastic density functional theory (SDFT) for the density field of the interacting particles. 
The SDFT  in question was first written down by Kawasaki \cite{kaw1994} on phenomenological grounds but  was later shown formally to be the equation of evolution for the density field of interacting Brownian particles \cite{dea1996}. The  SDFT  is analytically intractable when the particles interact, and very difficult to analyse even when there are no interactions \cite{vel2008}, however it can be used as the starting point for various approximate theories, for example mode coupling theory \cite{kim2014}. Recently it was shown that the linearised SDFT for charged particle systems yields a dynamical theory which has the Debye-Hückel theory as its static limit \cite{dea2014}. 
This has allowed the study of the dynamics of the thermal Casimir effect between adjacent plates containing Brownian charges. The same type of linearization approximation can also be used to 
compute the effective diffusion constant of a Brownian tracer particle in interaction with a bath of identical particles~\cite{dem2014}. Remarkably, the result obtained from this computation is identical to that obtained by an arduous one-loop renormalisation group analysis of the full $N$ particle Fokker-Planck equation~\cite{dea2004}. Finally we remark that the SDFT has even been used to derive integration results in random matrix theory~\cite{bun2014}. 

Here we will show how Onsager's results for electrolyte conductivity can be derived in a relatively straightforward fashion. 
As well as considering the case of a globally electroneutral electrolyte with two species of oppositely charged ions, we also deal with mobile ions that are not as an ensemble electroneutral. In this case, the mobile ion's total charge is neutralized by a uniform immobile background charge -- a so-called jellium model. We also generalize these results for the conductivity to lower dimensions $d$ with the 
corresponding $d$-dimensional Coulomb interaction. Finally, our results can be extended to systems with reduced dimension, for example ions confined to a two dimensional surface. 

\section{Model}\label{model}

We will consider a system containing two species of Brownian particles which interact via
pairwise interactions and which are acted on by a uniform external field. The basic formalism developed here can be applied to systems with an arbitrary number of species, however the reader will see below that on increasing the number of species the calculations become rather cumbersome. 
We will ultimately apply this model to the case of electrostatic interactions between charged particles in the presence of an applied field. The formalism developed however allows the possibility of including additional interactions between the ions. 
Let each species be denoted by an index $\alpha$ taking the label $+$ or $-$ depending if the ion is a cation or anion respectively. We denote by $\rho_\alpha({\xx},t)$ the local density operator of the species $\alpha$ and $\kappa_\alpha$ its mobility. We denote by $V_{\alpha\beta}$ the pairwise interaction between species $\alpha$ and $\beta$.  
We will ultimately decompose the pairwise interaction $V_{\alpha\beta}$ into an electrostatic and non-electrostatic contribution as
\begin{equation}\label{eq:general_int}
V_{\alpha\beta}({\xx}) = q^2 z_\alpha z_\beta G_0(\xx) + U_{\alpha\beta}(\xx).
\end{equation}
 Here $G_0(\xx)$ denotes the Green's function for the electrostatic interaction in the system, $q$ is the basic unit of charge and $z_\alpha$ denotes the valency of the ions of type $\alpha$. The interaction $U_{\alpha\beta}(\xx)$ denotes interactions other than the direct electrostatic interaction that may be present in the system, for example hard and soft core interactions. Note that in the case where there is a neutralizing uniform background charge (jellium model) there is no contribution to the electric field due to this uniform charge.

 At temperature $T$, the density fields obey the SDFT~\cite{dea1996}:
%\begin{equation}\label{eq:dean_initial}
%\partial_t\rho_\alpha = T\kappa_\alpha\nabla^2\rho_\alpha -\kappa_\alpha\nabla\cdot \left(\rho_\alpha \ff_\alpha \right) + \nabla\cdot \left[(\kappa_\alpha\rho_\alpha)^{1/2}\xxi_\alpha \right],
%\end{equation}
\begin{align}
\partial_t\rho_\alpha & = -\nabla\cdot \jj_\alpha, \label{eq:dean_conservation}\\
\jj_\alpha & = -T\kappa_\alpha\nabla\rho_\alpha+\kappa_\alpha\rho_\alpha \ff_\alpha+(\kappa_\alpha\rho_\alpha)^{1/2}\eeta_\alpha, \label{eq:dean_current}
\end{align}
where $\eeta_\alpha(\xx,t)$ is a Gaussian white noise with correlation function
\begin{equation}\label{eq:correl_noise}
\left\langle \eeta_\alpha(\xx,t)\eeta_\beta(\xx',t')\transp \right\rangle = 2T\delta_{\alpha\beta}\delta(\xx-\xx')\delta(t-t').
\end{equation}
In the above we have introduced the force $\ff_\alpha(\xx,t)$ generated by the applied external field $\EE$ and the interactions between the ions:
\begin{equation}\label{eq:elec_force}
\ff_\alpha = z_\alpha q \EE - \sum_\beta \nabla V_{\alpha\beta}*\rho_\beta,
\end{equation}
where $*$ denotes the convolution  over  spatial variables. 

The average electrical current is given by 
$\langle \JJ \rangle=q \sum_\alpha z_\alpha \langle \jj_\alpha \rangle$
and from this the conductivity $\sigma$ is defined via $ \langle \JJ \rangle=\sigma\EE$.
Inserting Eqs.~(\ref{eq:dean_current}) and (\ref{eq:elec_force}) in the definition of the current gives 
\begin{equation}
\langle \JJ \rangle=q^2\left(\sum_\alpha z_\alpha^2\kappa_\alpha\bar\rho_\alpha\right)\EE- q\sum_{\alpha,\beta} z_\alpha\kappa_\alpha \left\langle \rho_\alpha\nabla V_{\alpha\beta}*\rho_\beta \right\rangle.
\end{equation}
%The first term correspond to the current obtained when the fluctuations are neglected (see Eq.~\ref{condbare}).
To rewrite the second term, we introduce the density fluctuations and their stationnary equal time correlation as
\begin{align}
n_\alpha(\xx,t) & = \rho_\alpha(\xx,t)-\bar\rho_\alpha,\\
C_{\alpha\beta}(\xx) & = \langle n_\alpha(\xx,t) n_\beta(0,t) \rangle.
\end{align}
The average electric current is thus
\begin{equation}\label{eq:av_current_real_space}
\langle \JJ \rangle=\sigma_0\EE- q\sum_{\alpha,\beta} z_\alpha\kappa_\alpha \int C_{\alpha\beta}(\xx)\nabla V_{\alpha\beta}(\xx)\dd\xx.
\end{equation}
Neglecting the correlations between the ions, only the electric field contribution in the second term of Eq.~(\ref{eq:dean_current}) does not average to zero and the conductivity is given by Eq.~(\ref{condbare}). 
In order to compute the correction to the bare conductivity $\sigma_0$, we need to evaluate the correlations of the density fluctuations.

\section{General computation using the linearized Stochastic Density Functional Theory}\label{}

We have to use the SDFT (Eqs.~(\ref{eq:dean_conservation}-\ref{eq:elec_force})) to compute the stationnary correlation function $C_{\alpha\beta}(\xx)$. Unfortunately these equations are not linear and contain multiplicative noise,  thus the correlation function cannot be calculated in general.
This difficulty is circumvented by assuming small density variations, $n_\alpha(\xx,t)\ll\bar\rho_\alpha$, which allows one to linearize the SDFT~\cite{dea2014,dem2014}.
%\begin{equation}
%n_\alpha(\xx,t)=\rho_\alpha(\xx,t)-\overline \rho_\alpha \ll \overline \rho_\alpha.
%\end{equation}
The dynamics of $n_\alpha(\xx,t)$ become, to the lowest order
\begin{equation}
\partial_t n_\alpha=T\kappa_\alpha\nabla^2 n_\alpha - \kappa_\alpha z_\alpha q\EE\cdot\nabla n_\alpha + \kappa_\alpha \overline \rho_\alpha\nabla^2 \left[ \sum_\beta V_{\alpha\beta}*n_\beta\right]+(\kappa_\alpha\overline \rho_\alpha)^{1/2}\nabla\cdot\eeta_\alpha.
\end{equation}

The dynamics takes a simpler form in Fourier space, where $\tilde n_\alpha(\kk,t)=\int e^{-i\kk\cdot\xx} n_\alpha(\xx)\dd\xx$:
\begin{equation}\label{eq:dynamics_fourier}
\partial_t\tilde n_\alpha(\kk) = - \kappa_\alpha \left(T k^2 +iz_\alpha q\EE\cdot\kk \right)\tilde n_\alpha(\kk)-\kappa_\alpha  \bar\rho_\alpha k^2\sum_{\beta}\tilde V_{\alpha\beta}(\kk)\tilde n_\beta(\kk)+ \tilde \xi_\alpha(\kk,t),
\end{equation}
where the Gaussian noise $\tilde\xi_\alpha(\kk,t)$ has a correlation function
\begin{equation}
\left\langle \tilde\xi_\alpha(\kk,t)\tilde\xi_\beta(\kk',t') \right\rangle = 2(2\pi)^d T\kappa_\alpha \overline\rho_\alpha k^2\delta_{\alpha\beta}\delta(t-t')\delta(\kk+\kk').
\end{equation}

We can now rewrite the dynamics in terms of the two component vector 
\begin{equation}
N(\xx,t)= \begin{pmatrix}
n_+(\xx,t) \\ n_-(\xx,t)
\end{pmatrix},
\end{equation}
using the matrices
\begin{align}
\tilde R(\kk) & =  k^2 \begin{pmatrix}
\overline\rho_+\kappa_+ & 0 \\ 0 & \overline\rho_-\kappa_-
\end{pmatrix}, \label{eq:matrix_R}\\
\tilde A(\kk) &= T \begin{pmatrix}
\frac{1}{\overline\rho_+}\left(1+i \frac{z_+ q \EE\cdot\kk}{Tk^2} \right) + \frac{\tilde V_{++}(k)}{T} & \frac{\tilde V_{+-}(k)}{T} \\ \frac{\tilde V_{+-}(k)}{T} & \frac{1}{\overline\rho_-}\left(1+i \frac{z_- q \EE\cdot\kk}{Tk^2} \right) + \frac{\tilde V_{--}(k)}{T}
\end{pmatrix} \label{eq:matrix_A_Vab}.
\end{align}
The evolution Eq. (\ref{eq:dynamics_fourier}) now reads
\begin{equation}\label{eq:vector_density_fluctuations}
\partial_t\tilde N = -\tilde R\tilde A\tilde N + \tilde \Xi,
\end{equation}
where the correlation function of the Gaussian noise $\tilde \Xi(\kk,t)$ is given by
\begin{equation}
\left\langle \tilde \Xi(\kk,t)\tilde \Xi(\kk',t')\transp \right\rangle = 2(2\pi)^d T \tilde R(k) \delta(\kk+\kk')\delta(t-t').
\end{equation}

The density fluctuation correlation matrix $C_{\alpha\beta}(\xx)$ reads 
\begin{equation}
C(\xx)=\left\langle N(\xx) N(0)\transp \right\rangle,
\end{equation}
and  in  Fourier space, the correlation is given by
\begin{align}
\left\langle \tilde N(\kk)\tilde N(\kk')\transp \right\rangle &= \int e^{-i\kk\cdot\xx-i\kk'\cdot\xx'}C(\xx-\xx') \dd\xx\dd\xx'\\
&= \int e^{-i(\kk+\kk')\cdot\xx+i\kk'\cdot\uu}C(\uu)\dd\xx\dd\uu\\
& =(2\pi)^d\delta(\kk+\kk')\tilde C(\kk).
\end{align}
In the stationnary regime, $\tilde C$  satisfies~\cite{zwa2001}
%The time evolution of the equal-time correlation function is given by \cite{zwa2001}
%\begin{equation}\label{eq:evol_correl_function}
%\partial_t\tilde C = -\tilde R\tilde A\tilde C-\tilde C\tilde A^*\tilde R +2T\tilde R,
%\end{equation}
%In the stationnary regime, the equal-time correlation function is thus the solution of
\begin{equation}\label{eq:correlation_stationnary}
\tilde R\tilde A\tilde C+\tilde C\tilde A^*\tilde R =2T\tilde R.
\end{equation}
where we have used that $\tilde R(-\kk)\transp=\tilde R(\kk)$ and $\tilde A(-\kk)\transp=\tilde A(\kk)^*$. 
Without an electric field, the system would be in thermal equilibrium and $\tilde A$ would be self-adjoint; in this case one finds $\tilde C=T\tilde A^{-1}$, and one recovers the standard Debye-H\"uckel approximation for the density fluctuation correlation function. 
In the presence of an electric field, we have to solve the system of equations (\ref{eq:correlation_stationnary}).
The number of equations to solve is $M(M+1)/2$, where $M$ is the number of species; in our case of two species we have only three linear equations to solve.
%For the two species configuration considered here we see that one has a system of effectively three linear equations to solve, however on increasing the number of species the size of the system of equations becomes rapidly larger (scaling as $M^2$ if $M$ denotes the number of species). 

Writing $\rho_\pm\kappa_\pm = r_\pm$ and $\tilde A=T \begin{pmatrix} a & b \\ b & c \end{pmatrix}$, solving for the components of the correlation function leads to 
\begin{multline}\label{eq:correl_val}
\tilde C = \frac{2}{(a+a^*)(c+c^*)|r_+a+r_-c^*|^2-b^2 \left[r_+(a+a^*)+r_-(c+c^*) \right]^2}\times \\
\begin{pmatrix} (c+c^*)|r_+a+r_-c^*|^2 & -b (r_+a^*+r_-c) \left[r_+(a+a^*)+r_-(c+c^*) \right] \\ 
-b (r_+a+r_-c^*) \left[r_+(a+a^*)+r_-(c+c^*) \right] & (a+a^*)|r_+a+r_-c^*|^2
\end{pmatrix}.\end{multline}
Notice that the correlation function's off diagonal components satisfy  $\tilde C_{+-}(\kk) = \tilde C_{-+}^*(\kk)$, or equivalently 
$\tilde C_{+-}(\kk) = \tilde C_{-+}(-\kk)$, which in real space corresponds to $C_{+-}(\xx) = \ C_{-+}(-\xx)$, which is a symmetry condition pointed out by Onsager. This is because the symmetry $\xx\to -\xx$ is broken by the electric field (in the direction of the field), however reversing the direction of the field but at the same time swapping the charges generates a physically identical situation.

%The expression for the average electric current is given by
%\begin{equation}
%{\JJ} =  \left\langle\sum_{\alpha}  q z_\alpha \jj_\alpha\right\rangle= q\sum_{\alpha} \kappa_\alpha z_\alpha  \langle \rho_\alpha \ff_\alpha \rangle.
%\end{equation}
%The correction to the current with respect to the bare current is induced by the correlation between the density field $\rho_\alpha$ and the local force fields $\ff_\alpha$. In terms of the density fluctuation correlation matrix we thus find
%\begin{equation}
%\JJ = q^2 \left(\sum_\alpha z_\alpha^2\kappa_\alpha\rho_\alpha \right)\EE + q \sum_{\alpha,\beta}\kappa_\alpha z_\alpha \int i\kk \tilde V_{\alpha\beta}(-\kk) \tilde C_{\alpha\beta}(\kk) \frac{\dd\kk}{(2\pi)^d}.\label{ej}
%\end{equation}

The average electrical current, which is given by Eq.~(\ref{eq:av_current_real_space}), reads
\begin{equation}
\langle \JJ \rangle = \sigma_0\EE + q \sum_{\alpha,\beta}\kappa_\alpha z_\alpha \int i\kk \tilde V_{\alpha\beta}(-\kk) \tilde C_{\alpha\beta}(\kk) \frac{\dd\kk}{(2\pi)^d}.\label{ej}
\end{equation}
Recall here that the correction to the bare current is given by the average of the interaction term which was neglected upon linearizing the full SDFT, thus our expansion is only valid when the computed correction is small.
From the expression for the current we can define the field dependent conductivity via
\begin{equation}
\langle \JJ \rangle = \sigma(E) \EE = [\sigma_0+\Delta\sigma(E)]\EE,
\end{equation}
where we assumed that the system is isotropic.
%Due to the isotropy of the system $\sigma = \sigma(E)$ is only a function of the strength $E$  of the applied field.
%If we define  
%\begin{equation}
%\sigma(E) = \sigma_0 + \Delta\sigma(E),
%\end{equation}
We find that the correction to the bare conductivity is given by
\begin{multline}
\Delta\sigma(E) = -\frac{q^2\overline\rho_+\overline\rho_-(\kappa_+z_+-\kappa_- z_-)^2}{(\kappa_++\kappa_-)T^2}\\
\times\int 
\frac{\frac{k_\parallel^2}{k^2} \tilde V_{+-}^2 \left(1+\frac{\kappa_+\overline\rho_+\tilde V_{++}+\kappa_-\overline\rho_-\tilde V_{--}}{(\kappa_++\kappa_-)T} \right)}
{\begin{array}{c} \left(1+\frac{\overline\rho_+\tilde V_{++}}{T} \right) \left(1+\frac{\overline\rho_-\tilde V_{--}}{T} \right) \left(\left[1+\frac{\kappa_+\overline\rho_+\tilde V_{++}+\kappa_-\overline\rho_-\tilde V_{--}}{(\kappa_++\kappa_-)T} \right]^2 + \left[\frac{(\kappa_+ z_+ - \kappa_-z_-)qEk_\parallel}{(\kappa_++\kappa_-)Tk^2} \right]^2 \right) \\
-\frac{\overline\rho_+\overline\rho_-\tilde V_{+-}^2}{T^2}\left(1+\frac{\kappa_+\overline\rho_+\tilde V_{++}+\kappa_-\overline\rho_-\tilde V_{--}}{(\kappa_++\kappa_-)T} \right)^2 \end{array}
 }\frac{\dd\kk}{(2\pi)^d},\label{gensig}
\end{multline}
where $k_\parallel$ denotes the component of the vector $\kk$ in the direction of $\EE$ (the third line of the equation belongs to the denominator of the integrand).

We notice from Eq.~(\ref{gensig}) that the correction $\Delta\sigma(E)$ is zero when $\kappa_+z_+ = \kappa_- z_-$. In this case, %at the first order level of approximation, 
the two ionic types move with the same average velocity $\vv = q\kappa_\pm z_\pm \EE$ due to an applied uniform field and the density fluctuation correlation keeps its equilibrium form, which is isotropic and does not modify the average electric field felt by an ion.

At zero field, the correction to the conductivity is given by
\begin{multline}\label{eq:general_nofield}
\Delta\sigma(0)=-\frac{q^2\overline\rho_+\overline\rho_-(\kappa_+z_+-\kappa_- z_-)^2}{d(\kappa_++\kappa_-)T^2}\\\times\int \frac{\tilde V_{+-}(k)^2}{\left[1+\frac{\overline\rho_+\tilde V_{++}(k)+\overline\rho_-\tilde V_{--}(k)}{T}+\frac{\overline\rho_+\overline\rho_-\left(\tilde V_{++}(k)\tilde V_{--}(k)-\tilde V_{+-}(k)^2 \right)}{T^2} \right] \left[1+\frac{\kappa_+\overline\rho_+\tilde V_{++}(k)+\kappa_-\overline\rho_-\tilde V_{--}(k)}{(\kappa_++\kappa_-)T} \right]}\frac{\dd\kk}{(2\pi)^d}.
\end{multline}

\section{Zero field conductivity for purely electrostatic interactions}

\subsection{General results for a purely electrostatic interaction}\label{}

If the interaction is purely electrostatic, i.e., only the first term is present in Eq.~(\ref{eq:general_int}), our result at zero field Eq.~(\ref{eq:general_nofield}) reduces to
\begin{equation}\label{eq:electro_nofield}
\Delta\sigma(0) =-\frac{q^6\bar\rho_+\bar\rho_- z_+^2z_-^2(\kappa_+z_+-\kappa_-z_-)^2}{d(\kappa_++\kappa_-)T^2}\int \frac{\tilde G_0(\kk)^2}{\left(1+\frac{q^2(z_+^2\bar\rho_++z_-^2\bar\rho_-)\tilde G_0(\kk)}{T} \right) \left(1+\frac{\sigma_0\tilde G_0(\kk)}{(\kappa_++\kappa_-)T} \right)}\frac{\dd\kk}{(2\pi)^d}.
\end{equation}
At this point we should note that theories with a $\kk$ dependent (non-local in space) dielectric function \cite{bop1996} can also be treated  within this formalism.
We now define, following the notation of Onsager~\cite{ons1927},
\begin{align}
m^2_\pm &= \frac{\overline \rho_\pm z_\pm^2 q^2}{\epsilon T} \\
m^2 &= m_+^2 + m_-^2 \\
m'^2 &= \frac{\kappa_+ m_+^2 +\kappa_- m_-^2}{\kappa_++\kappa_-}= \frac{\sigma_0}{\epsilon T(\kappa_+ +\kappa_-)}.
\end{align}
Using this, we can write
\begin{equation}\label{eq:electro_nofield_masses}
\frac{\Delta\sigma(0)}{\sigma_0}=-\frac{q^2}{d\epsilon T}\frac{m_+^2m_-^2(\kappa_+z_+-\kappa_-z_-)^2}{m'^2(\kappa_++\kappa_-)^2} 
%I\left[\epsilon\tilde G(\kk),m,m'\right]
\int \frac{1}{\left([\epsilon\tilde G_0(\kk) ]^{-1}+m^2 \right) \left([\epsilon\tilde G_0(\kk) ]^{-1}+m'^2 \right)}\frac{\dd\kk}{(2\pi)^d}.
\end{equation}

In the case where all the charges are mobile in an electroneutral system, $\sum_\alpha z_\alpha \bar\rho_\alpha = 0$ and thus
\begin{equation}
\frac{m_+^2 m_-^2 (\kappa_+z_+-\kappa_-z_-)^2}{m'^2 (\kappa_+ +\kappa_-)^2}= 
-z_+ z_- m'^2,
\end{equation}
leading to a simpler form for the correction:
\begin{equation}\label{eq:electro_nofield_neutral}
\frac{\Delta\sigma(0)}{\sigma_0}=\frac{q^2 z_+ z_- m'^2}{d\epsilon T}
\int \frac{1}{\left([\epsilon\tilde G_0(\kk) ]^{-1}+m^2 \right) \left([\epsilon\tilde G_0(\kk) ]^{-1}+m'^2 \right)}
\frac{\dd\kk}{(2\pi)^d}.
\end{equation}

\subsection{Conductivity for ions moving in homogeneous space}\label{}

In a homogeneous $d$-dimensional space with solvent dielectric permittivity $\epsilon$, the electrostatic Green function is given by $\tilde G_0(\kk)=(\epsilon k^2)^{-1}$ and the conductivity correction (Eq.~(\ref{eq:electro_nofield_masses})) becomes
\begin{equation}\label{eq:conductivity_diff_val_int}
\frac{\Delta\sigma(0)}{\sigma_0}=-\frac{q^2}{d \epsilon T}\frac{m_+^2 m_-^2 (\kappa_+z_+-\kappa_-z_-)^2}{m'^2 (\kappa_+ +\kappa_-)^2}\int \frac{1}{(k^2+m^2)(k^2+m'^2)}\frac{\dd\kk}{(2\pi)^d}.
\end{equation}
%where, following the notation of Onsager \cite{ons1927} 
%\begin{eqnarray}
%m^2_\pm &=& \frac{\overline \rho_\pm z_\pm^2 q^2}{\epsilon T} \\
%m^2 &=& m_+^2 + m_-^2 \\
%m'^2 &=& \frac{\kappa_+ m_+^2 +\kappa_- m_-^2}{\kappa_++\kappa_-}= \frac{\sigma_0}{\epsilon T(\kappa_+ +\kappa_-)}
%\end{eqnarray}
We note that while $m$ is the usual inverse Debye screening length, which only depends on static quantities, the inverse length scale $m'$ depends generally explicitly on the dynamical properties of the system via the mobilities $\kappa_\alpha$. 
%However, in the case where the electrolyte is monovalent we see that $m' = m/\sqrt{2}$ and is thus independent of the ionic mobilities. 
However, for a monovalent electrolyte ($z_\pm=\pm 1$), $m' = m/\sqrt{2}$ and is thus independent of the ionic mobilities. 
Furthermore, we note that the correction to the bare conductivity is always negative.
Explicit evaluation for dimensions $d=1$ to $3$ yields
\begin{align}
\left(\frac{\Delta\sigma(0)}{\sigma_0} \right)_{d=3} & =-\frac{q^2}{12\pi \epsilon T}\frac{m_+^2 m_-^2 (\kappa_+z_+-\kappa_-z_-)^2}{m'^2(\kappa_+ +\kappa_-)^2}\frac{1}{m+m'},\\
\left(\frac{\Delta\sigma(0)}{\sigma_0} \right)_{d=2} & = -\frac{q^2}{4\pi\epsilon T}\frac{m_+^2 m_-^2 (\kappa_+z_+-\kappa_-z_-)^2}{m'^2 (\kappa_+ +\kappa_-)^2}\frac{\log \left(\frac{m}{m'} \right)}{m^2-m'^2},\\
\left(\frac{\Delta\sigma(0)}{\sigma_0} \right)_{d=1} & = -\frac{q^2}{2\epsilon T}
\frac{m_+^2 m_-^2 (\kappa_+z_+-\kappa_-z_-)^2}{m'^2 (\kappa_+ +\kappa_-)^2}
\frac{1}{mm'(m+m')}.\\
\end{align}

%In the case where all the charges are mobile in an electroneutral system, $\sum_\alpha z_\alpha \bar\rho_\alpha = 0$ and thus
%\begin{equation}
%\frac{m_+^2 m_-^2 (\kappa_+z_+-\kappa_-z_-)^2}{m'^2 (\kappa_+ +\kappa_-)^2}= 
%-z_+ z_- m'^2,
%\end{equation}
For an electroneutral system where all charges are mobile, the correction is given by
\begin{align}
\left(\frac{\Delta\sigma}{\sigma_0} \right)_{d=3} & = \frac{q^2z_+ z_-}{12\pi\epsilon T}\frac{m'^2}{m+m'},\\
\left(\frac{\Delta\sigma}{\sigma_0} \right)_{d=2} & = \frac{q^2z_+ z_-}{4\pi\epsilon T}\frac{\log \left(\frac{m}{m'} \right)}{\left(\frac{m}{m'} \right)^2-1},\\
\left(\frac{\Delta\sigma}{\sigma_0} \right)_{d=1} & = \frac{q^2z_+ z_-}{2\epsilon T}\frac{m'}{m(m+m')}.\\
\end{align}
The result in dimension three agrees with that obtained by Onsager \cite{ons1927}.

We now recall that the validity of the linearization of the SDFT can be checked a posteriori be verifying that the correction to the average current or conductivity is small.  
In all dimensions this condition is satisfied if the electrostatic coupling constant $\Gamma =\frac{q^2z_+ z_-}{2\pi\epsilon T}$ is small, but the dependence on the density depends on the dimension. For example, for monovalent ions:
\begin{itemize}
\item For $d=3$, the correction is proportional to $\sqrt{\bar\rho}$ and is small if $\bar\rho$ is small.
\item For $d=2$, the magnitude of the correction is purely controlled by the coupling constant $\Gamma$, which is dimensionless here, and does not depend on the density. In $d=2$ it is well known that there is a transition from a weak coupling conducting phase to a strong coupling dielectric phase (where the charge carriers exist only in bound pairs) -- the Kosterlitz-Thouless transition~\cite{kos1973}. The weak coupling approach applied here is clearly only valid in the conducting phase.
\item For $d=1$, the correction is proportional to $1/\sqrt{\bar\rho}$ and is thus small at high densities, however the long range nature of the electric field generated by one dimensional charges means that the one dimensional problem is not very realistic. 
\end{itemize}
%For monovalent ions we see that the same  condition can be achieved in three dimensions by taking $\sqrt{\overline\rho}$ small as here the correction term is proportional to ${\overline\rho}$.
%In two dimensions however, again for monovalent systems, the correction is purely controlled by the, here dimensionless coupling constant $\Gamma$, and is independent of the density. In two dimensions it is well known that there  is a transition from a weak coupling conducting phase to a strong coupling dielectric phase (where the charge carriers exist in bound pairs)- the Kosterlitz-Thouless transition. The weak coupling  approach applied here is clearly only valid in the conducting phase. 
%In one dimension the correction decays as $1/\sqrt{\rho}$ and the correction is thus only small at high densities, however the long range nature of the electric field generated by one dimensional charges means that the one dimensional problem is not very realistic. 

If, instead of the Coulomb interaction, the particles interact with a Yukawa or screened Coulomb interaction, with screening length $\xi$ ({\em i.e} the relevant Green's function becomes $\tilde G_0(\kk) = [\epsilon(k^2+\xi^{-2})]^{-1}$),
in our precedent calculations this simply amounts to replacing the inverse Debye length $m$ by $\sqrt{m^2+\xi^{-2}}$, and likewise for  $m'$, in the integral over $\kk$ in Eq. (\ref{eq:conductivity_diff_val_int}). 
%\red{ mais les  prefacteurs devant ne changent pas non ? Donc je pense pas que les formules sature dans la limite ou rho ->0}

\subsection{Conductivity for ions confined to $d=1$ or $2$ and interacting with the 3d Green function}\label{}

We now turn to the case of charges constrained to move in a low dimensional space ($d=1$ or $2$) but that still interact with the 3-dimensional interaction $G_0(\xx)=1/(4\pi\epsilon |\xx|)$. 
In this case we find that the expressions for the conductivity corrections for systems confined in both two and one dimensions exhibit ultraviolet (large $k$) divergences. These divergences arise due to the singular nature of the three dimensional Coulomb interaction at short distances.
The Coulomb  interaction between the charges can be regularized by using a Gaussian distribution of charge about each ion type to give a local charge density
\begin{equation}
\rho_{c\alpha}(\xx) = \frac{q z_\alpha}{(2\pi \ell^2_\alpha)^{d\over 2}} \exp\left(-\frac{\xx^2}{2\ell_\alpha^2}\right),
\end{equation}
where $\ell_\alpha$ is the size of the region on the ion over which the net charge is localized. In Fourier space this gives an effective electrostatic interaction
\begin{equation}
\tilde V_{\alpha\beta}(\kk) = \frac{q^2z_\alpha z_\beta}{\epsilon k^2}\exp\left(-\frac{k^2[\ell^2_\alpha + \ell^2_\beta]}{2}\right).
%\tilde G_0(\kk) =  \frac{\exp\left[-\frac{k^2(\ell^2_\alpha + \ell^2_\beta)}{2}\right]}{\epsilon k^2}.
\end{equation}

\subsubsection{Confinement to a plane.}\label{}

Now if we consider a 3 dimensional interaction between charges restricted to the plane $z=0$, the two dimensional Fourier transform of the effective electrostatic interaction in the $(x,y)$ plane is given by
\begin{equation}
\tilde V_{\alpha\beta}^{d=2}(\kk) = \int \tilde V_{\alpha\beta}(\kk, k_z) \frac{\dd k_z }{2\pi} 
%\tilde G_0^{(2d)}({\kk}) = \frac{1}{2\pi} \int \tilde G_0(\kk, k_z) \dd k_z 
\end{equation}
where here $\kk$ is a two dimensional vector. This then gives
\begin{equation}
%\tilde G_0^{(2d)}
\tilde V_{\alpha\beta}^{d=2}({\kk}) = \frac{q^2z_\alpha z_\beta}{\pi\epsilon k} \int_0^\infty  
\frac{\exp\left(-\frac{[\ell^2_\alpha + \ell^2_\beta]k^2[u^2+1]}{2}\right)}{u^2+1} \dd u.
\end{equation}
In two dimensions on taking the limit $\ell_\alpha \to 0$ for both ions we recover the standard result $\tilde G_0^{d=2}(\kk)= 1/(2\epsilon k)$; clearly this same behavior is recovered in the small $k$ limit and the long distance properties of the Coulomb interaction are thus unaffected by the short distance regularization. 
For large $k$,
\begin{equation}
\tilde V_{\alpha\beta}^{d=2}({\kk}) \underset{k\to\infty}{\sim} \frac{q^2z_\alpha z_\beta}{\sqrt{2\pi}\epsilon k^2\sqrt{\ell^2_\alpha + \ell^2_\beta}}\exp\left(-\frac{k^2[\ell^2_\alpha + \ell^2_\beta]}{2}\right);
\end{equation}
subsequently all the formula for the conductivities derived in two dimensions are now rendered finite. 

For simplicity if we take $\ell_+ = \ell_-=\ell$ and define $s = \sqrt{2}\ell$, the interaction potential can be written in the scaling form
\begin{equation}
\tilde V_{\alpha\beta}^{d=2}({\kk}) = \frac{q^2z_\alpha z_\beta}{\epsilon}sf(ks).
\end{equation}
The resulting integrals for the conductivity correction cannot be carried out analytically but the leading divergence as $s\to 0$ can be extracted by writing  $\tilde V^{d=2}_{\alpha\beta}(\kk)= q^2 z_\alpha z_\beta/(2\epsilon k)$ for $ks \ll 1$ and cutting off the resulting integral at $k=1/s$; the regularisation is thus equivalent to the Pauli-Villars regularisation~\cite{Itzykson2006}.
The result of the regularisation is that the correction to the bare conductivity behaves as 
\begin{equation}
\left(\frac{\Delta\sigma(0)}{\sigma_0}\right)_{d=2,\,\textrm{3d int.}} \underset{s\to 0}{\sim} -\frac{q^2 z_+z_- m'^2 \log(sm^2)}{16\pi\epsilon T}.
\end{equation}
%We thus see that t
The prefactor to the logarithmic correction is small at small coupling constant $\Gamma$ and at small density; the argument of the logarithmic term behaves as
$s \Gamma \overline \rho$.  For charges restricted to a plane we thus see that the
change in the conductivity due to interactions behaves as $ \overline \rho\log(\Gamma \overline \rho s)$ as opposed to the $\sqrt{\overline \rho}$ correction seen when they are free to move in three dimensions.

\subsubsection{Confinement to a line.}\label{}

In one dimension, for charges restricted to the line $y=z=0$, the effective interaction is
\begin{equation}
\tilde V_{\alpha\beta}^{d=1}(k)  = \int \tilde V_{\alpha\beta}(k,k_y,k_z)  \frac{\dd k_z \dd k_y}{{(2\pi)^2}},
\end{equation}
and thus
\begin{equation}
%\tilde G_0^{(1d)}(k) = 
\tilde V_{\alpha\beta}^{d=1}(k)=\frac{q^2 z_\alpha z_\beta}{2\pi\epsilon} \int_0^\infty \frac{\exp\left(-\frac{[\ell^2_\alpha + \ell^2_\beta]k^2 [u^2+1]}{2}\right)}{u^2+1} u \dd u.
\end{equation}
The asymptotic behavior of the potential is given by
\begin{align}
\tilde V_{\alpha\beta}^{d=1}(k) & \underset{k\to 0}{\sim} -\frac{q^2 z_\alpha z_\beta}{2\pi\epsilon}\log\left(k\sqrt{\ell^2_\alpha + \ell^2_\beta}\right),\\
\tilde V_{\alpha\beta}^{d=1}(k) & \underset{k\to\infty}{\sim} \frac{1}{2\pi \epsilon k^2 (\ell^2_\alpha + \ell^2_\beta)}\exp\left(-\frac{k^2 [\ell^2_\alpha + \ell^2_\beta]}{2}\right).
\end{align}
Again this regularises all integrals appearing in the formulas for the conductivity both at zero and finite field.

% In the limit $k\to 0$ we find that
%\begin{equation}
%\tilde V_{\alpha\beta}^{d=1}(k)\underset{k\to 0}{\sim} -\frac{q^2 z_\alpha z_\beta}{2\pi\epsilon}\log\left(k\sqrt{\ell^2_\alpha + \ell^2_\beta}\right),
%\end{equation}
%while as $k\to \infty$ one finds
%\begin{equation}
%\tilde V_{\alpha\beta}^{d=1}(k)\underset{k\to\infty}{\sim} \frac{1}{2\pi \epsilon k^2 (
%\ell^2_\alpha + \ell^2_\beta)}\exp\left(-\frac{k^2 [\ell^2_\alpha + \ell^2_\beta]}{2}\right).
%\end{equation}

In the simple case where $\ell_+=\ell_-=s/\sqrt{2}$, the scaling form of $\tilde V_{\alpha\beta}^{d=1}(k)$ is given by
\begin{equation}
\tilde V_{\alpha\beta}^{d=1}(k) = \frac{q^2 z_\alpha z_\beta}{\epsilon}f(ks).
\end{equation}
% the result of the regularization is a correction to the bare conductivity behaves as 
%and the 
The regularized correction to the bare conductivity thus behaves as
\begin{equation}
\left(\frac{\Delta\sigma(0)}{\sigma_0}\right)_{d=1,\,\textrm{3d int.}} \underset{s\to 0}{\sim}  -\frac{q^2 z_+z_- m'^2}{\pi\epsilon T s}g(m,m')
\end{equation}
where $g(m,m')$ is a finite function of $m$ and $m'$ given by
\begin{equation}
g(m,m') = \int_0^\infty \frac{\dd p}{[f(p)^{-1} + m^2][f(p)^{-1} + m'^2]}.
\end{equation}
Interestingly, the integral defining $g$ turns out to be convergent when $m=m'=0$. This means that, for small densities, the change in the conductivity due to interactions  is proportional to $\overline \rho$.

\section{Field dependence for monovalent salts}

\subsection{Correlation function}\label{}

\begin{figure}
\begin{center}
\includegraphics{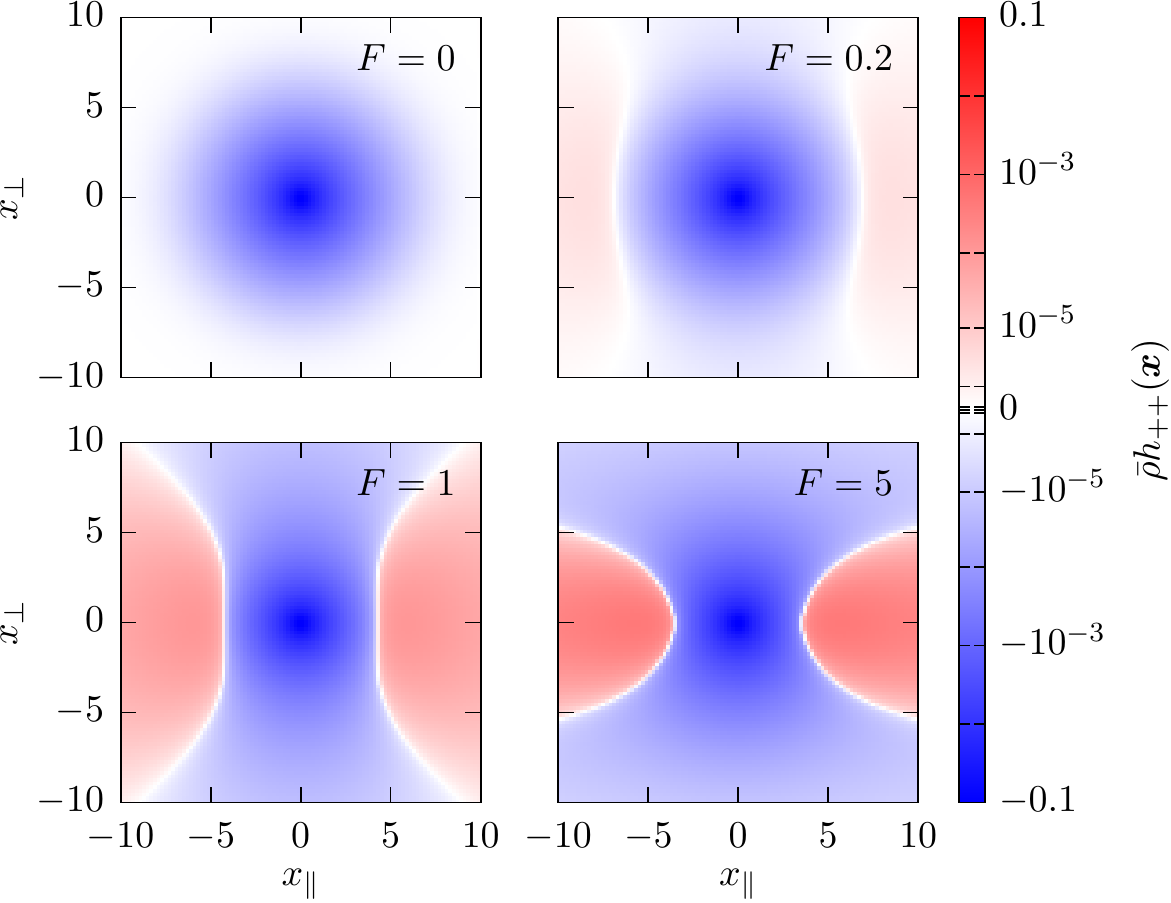}
\end{center}
\caption{Pair correlation function $h_{++}(\xx)$ (Eq.~(\ref{eq:correlation_function})) for different values of the dimensionless electric field $F=qE/(mT)$ (Eq.~(\ref{eq:dimensionless_field})); the unit of length is the Debye screening length $m^{-1}$ defined in Eq.~(\ref{eq:m}). 
The coordinates $x_\parallel$ and $x_\perp$ denote the directions parallel and perpendicular, respectively, to the field direction. 
Multiplied by $\bar\rho$, the pair correlation function does not depend on $\bar\rho$. As explained in the main text, $h_{++}(\xx)$ is regularized by convolution with a Gaussian whose width is set to the pixel size.
}
\label{fig:cpp_f}
\end{figure}

\begin{figure}
\begin{center}
\includegraphics{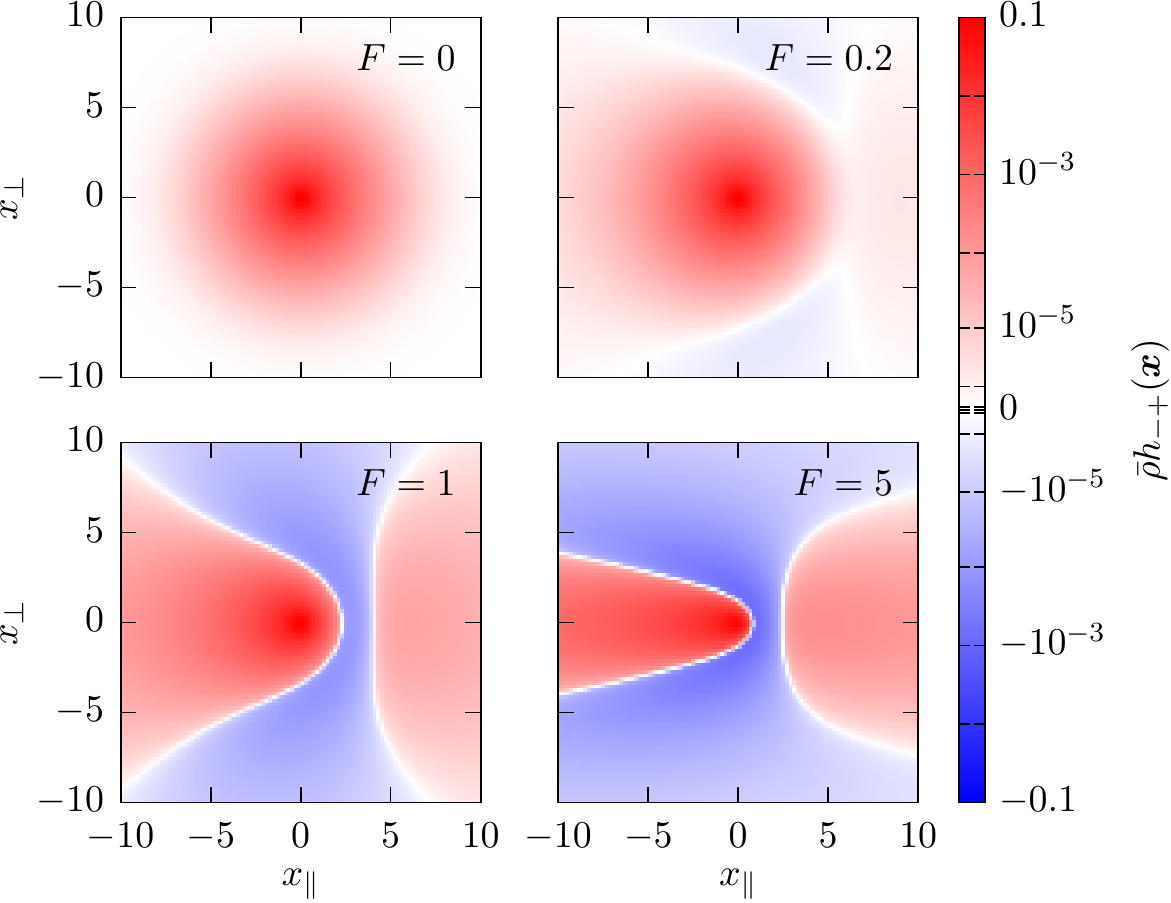}
\end{center}
\caption{Pair correlation function $h_{-+}(\xx)$ (Eq.~(\ref{eq:correlation_function})) in monovalent electrolytes for different values of the dimensionless electric field $F=qE/(mT)$ (Eq.~(\ref{eq:dimensionless_field})); the unit of length is the Debye screening length $m^{-1}$ (Eq.~(\ref{eq:m})). The coordinates $x_\parallel$ and $x_\perp$ denote the directions parallel and perpendicular, respectively, to the field direction.
Multiplied by $\bar\rho$, the pair correlation function does not depend on $\bar\rho$. As explained in the main text, $h_{-+}(\xx)$ is regularized by convolution with a Gaussian whose width is set to the pixel size.}
\label{fig:cmp_f}
\end{figure}

Here we examine the case of monovalent salts (with no background charge, $z_+=-z_-=1$ and $\bar\rho_+=\bar\rho_-=\bar\rho$), which turns out to be the simplest case where one can obtain completely analytical results for the conductivity at any applied field. We define the inverse Debye length $m$
and dimensionless electric field $\FF$ by
\begin{align}
m^2 & = \frac{2\overline\rho q^2}{\epsilon T}, \label{eq:m}\\
\FF & = \frac{q}{mT}\EE. \label{eq:dimensionless_field}
\end{align}
%and as the system in electroneutral we have written $\overline \rho_\pm= \overline \rho$.
For this case the correlation function is given by
\begin{equation}\label{eq:correlation_function}
\tilde C(\kk) = \frac{\overline\rho}{\left(1+\frac{m^2}{2k^2} \right) \left[1+\frac{m^2}{k^2}+\left(\frac{m\kk\cdot\FF}{k^2} \right)^2 \right]} \begin{pmatrix}
\left(1+\frac{m^2}{2k^2} \right)^2+\left(\frac{m\kk\cdot\FF}{k^2} \right)^2 & \frac{m^2}{2k^2}\left(1+\frac{m^2}{2k^2}-i \frac{m\kk\cdot\FF}{k^2} \right)\\ \frac{m^2}{2k^2}\left(1+\frac{m^2}{2k^2}+i \frac{m\kk\cdot\FF}{k^2} \right) & \left(1+\frac{m^2}{2k^2} \right)^2+\left(\frac{m\kk\cdot\FF}{k^2} \right)^2
\end{pmatrix}
\end{equation}
The pair correlation functions $h_{++}(\xx)$ and $h_{-+}(\xx)$, defined by
\begin{equation}
h_{\alpha\beta}(\xx) = \frac{1}{\overline\rho^2} C_{\alpha\beta}({\xx}) -\frac{\delta_{\alpha\beta}\delta(\xx)}{\overline{\rho}},
\end{equation}
are plotted in Figs.~\ref{fig:cpp_f} and~\ref{fig:cmp_f} for different values of the electric field; they give the change in density of cations and anions, respectively, around a cation. 
These figures are obtained by inversing the Fourier transforms $\tilde h_{\alpha\beta}(k)$. 
Despite the fact that the expressions for the conductivity that we have found are finite, the 
integral of $\tilde h_{\alpha\beta}(k)$ is divergent at large $k$, meaning that the inverse Fourier transform needs to be regularized (the corresponding integrals at large $k$ decay as $1/k^2$ indicating a $1/|\xx|$ divergence at small $|\xx|$ in three dimensions). 
We have thus  regularized the resulting integrals  using a Gaussian cut-off function $\exp(-\ell^2 k^2/2)$, the pair correlation function shown is thus the convolution of the corresponding  pair correlation functions with a Gaussian of width $\ell$.
In Figs.~\ref{fig:cpp_f} and \ref{fig:cmp_f}, the cut-off length is set to the pixel size, $\ell=0.2m^{-1}$.

In Fig.~\ref{fig:cpp_f}, at zero field, $h_{++}(\xx)$ is negative about the origin and then decays monotonically to zero at large distances. When $F\neq 0$ we see that the correlations have a longer range as screening is destroyed by the field.  At larger values of $F$, the probability of an other positive ion being present around a given positive ion is largest in the direction of the field as indicated by the lobes on the figure. 
In Fig.~\ref{fig:cmp_f}, without external field, the correlation function is positive about the origin and decays to zero as the distance from the origin increases, indicating screening of positive charges (at the origin) by negative charges. As $F$ is turned on, the negative ions are again more likely to be found in the direction of the applied field but most likely in front of the positive charge than behind, which of course makes physical sense. 
There is also a region of depletion of negative ions further from the cations, an effect which as also been seen in the pair correlation function in recent simulations of a lattice based ionic model \cite{kai2013} (the correlation function $g_{-+}(\xx)=h_{-+}(\xx)+1$ becomes less than 1 in their Fig.~4). 
%At very large values of $F$ the positive ions in front of the negative charge become much more concentrated in the direction of the field than those behind. 
Both figures thus indicate a tendency for the positive and negative ions to form chains in the direction of the field. 
%A similar increase in the probability to find the positive ions in the regions afore and astern a negative ion have also been seen in the pair correlation function in recent simulations of a lattice based ionic model \cite{kai2013} (their  Fig.~4). 

\subsection{Conductivity}\label{}

\begin{figure}
\begin{center}
\includegraphics{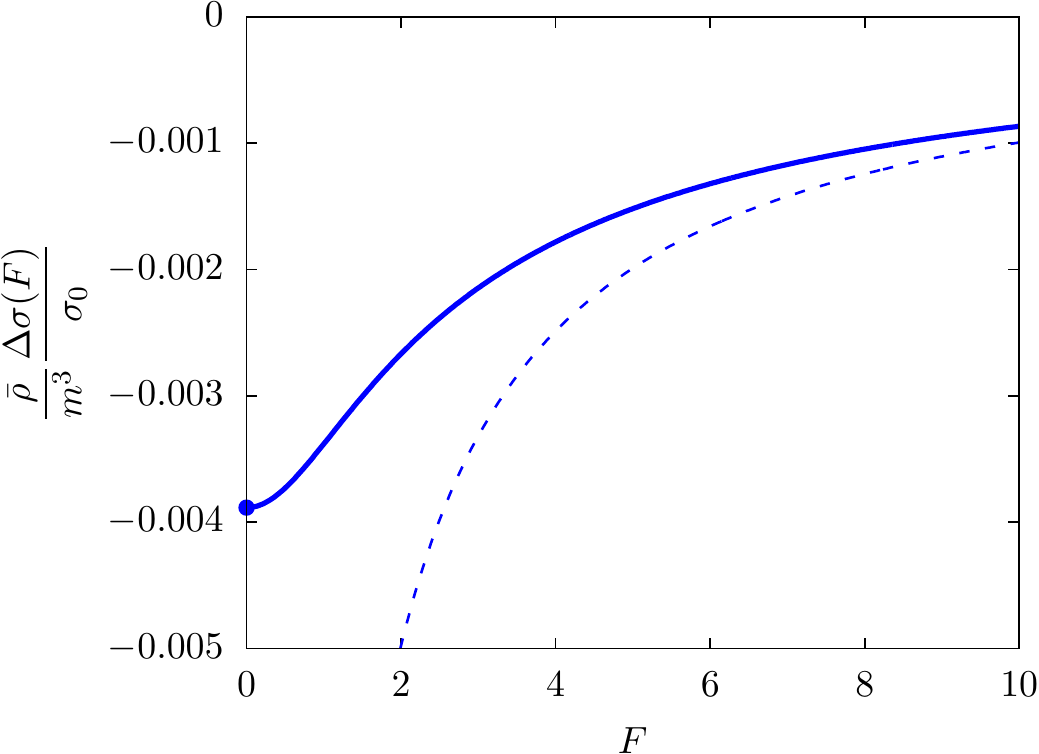}
\end{center}
\caption{Conductivity correction for $d=3$ as a function of the (dimensionless) applied electric field (solid line), and asymptotic expression $\Delta\sigma\sim 1/F$ (dashed).}
\label{fig:f_corr}
\end{figure}

The field dependent conductivity 
is given by
\begin{equation}
\frac{\sigma(F)}{\sigma_0}=1-\frac{m^d}{2\overline\rho}\int\frac{u_\parallel^2}{(2u^2+1)(u^4+u^2+F^2u_\parallel^2)}\frac{\dd\uu}{(2\pi)^d},
\end{equation}
where, again,   the subscript ``$\parallel\,$'' denotes the component parallel to the electric field $\EE$.
For a finite field and a dimension $d=3$, we need to evaluate the integral
\begin{equation}
I=\int\frac{u_\parallel^2}{(2u^2+1)(u^4+u^2+F^2u_\parallel^2)}\frac{\dd\uu}{(2\pi)^3} = \frac{1}{2\pi^2}\int_0^\infty \dd u \int_0^1 \dd v \frac{u^2v^2}{(2u^2+1)(u^2+1+F^2v^2)}.
\end{equation}
The integral over $u$ gives
\begin{align}
I & = \frac{1}{8\pi}\int_0^1 \frac{2(1+F^2v^2)-\sqrt{2}\sqrt{1+F^2v^2}}{(1+2F^2v^2)\sqrt{1+F^2v^2}} v^2\dd v\\
& = \frac{1}{16\pi F^3}\left[F\sqrt{1+F^2}-\arctan \left(\frac{F}{\sqrt{1+F^2}} \right)-\sqrt{2}F+\arctan(\sqrt{2}F)\right].
\end{align}
This gives the final result
\begin{equation}\label{eq:corr_finite_field}
\frac{\sigma(F)}{\sigma_0}=1-\frac{m^3}{32\pi \overline\rho F^3}\left[F\sqrt{1+F^2}-\arctan \left(\frac{F}{\sqrt{1+F^2}} \right)-\sqrt{2}F+\arctan\left(\sqrt{2}F\right)\right].
\end{equation}
The first term in brackets is dominant at large fields and the correction decays as $1/F$. This result agrees with that of Onsager and Kim \cite{ons1957}. The expression Eq. (\ref{eq:corr_finite_field}) is plotted on Fig.~\ref{fig:f_corr}. The correction decays as the applied field increases, which is the so called Wien effect in simple strong electrolytes.

In lower dimensions, with the corresponding Coulomb interaction in that dimension, the conductivity at finite field is given by
\begin{align}
\left(\frac{\sigma(F)}{\sigma_0} \right)_{d=1} & = 1-\frac{m}{4\overline\rho}\frac{1}{\sqrt{F^2+1}\left(\sqrt{2(F^2+1)}+1 \right)},\\
\left(\frac{\sigma(F)}{\sigma_0} \right)_{d=2} & = 1 - \frac{m^2}{16\pi\overline\rho}\left(\frac{1}{F^2 \sqrt{1+2 F^2}} \left[\log\left(1+F^2+\sqrt{1+2 F^2}\right) \right.\right. \nonumber \\
 & \qquad \left. - \log\left(2+3F^2 + 2\sqrt{(1+F^2) (1+2 F^2)}\right)\right] \\
 & \qquad \left.+ \frac{1}{F^2}\log\left(1+\frac{F^2}{2}+\sqrt{1+F^2}\right) \right).\nonumber 
\end{align}

\section{Conclusion}  
We have used SDFT to compute the conductivity of strong electrolytes. Our approach
is rather simple and compact and reproduces classical results of Onsager and collaborators which were derived by very different methods. We have furthermore seen how the calculations of Onsager are modified when there is a static uniform background charge. In this situation, even a system where all ions have the same charge but have different mobilities exhibits a  negative correction to  the bare conductivity. 
Our formalism has also allowed us to generalize the results of Onsager to Coulomb systems in arbitrary dimensions. 

Clearly our results are applicable to more general systems; they can be applied to more complex electrostatic models, for instance models with non-local dielectric constants \cite{bop1996}. 
%The general formulas given here can also be applied to the motion of ions confined to planar surfaces and in this context can be used to take into account the effect of image charges due to dielectric discontinuities. 
The general formulas given here can also be applied to the motion of ions confined to quasi-1d geometries, e.g. in carbon or boron-nitride nanotubes~\cite{Liu2010,Siria2013,Secchi2015}.
%Finally general non-electrostatic interactions can also be treated within the formalism, 
Finally, our formalism can treat general non-electrostatic interactions like steric interactions, which enters in the dynamics of ionic liquids~\cite{Lee2015}.

Further extensions of this work are clearly possible, one could for example analyze the influence of  a solvent on the conductivity. The effect of a solvent in Onsager's calculations is in fact additive at the level of approximation he employed, the additional term is generated by the solvent induced flow due to the movement of the ions. Using the formalism here 
it is possible that the perturbation analysis could be carried out in a self consistent manner thus potentially improving its results at higher concentrations in three dimensions.

\section*{References}

%\bibliographystyle{iopart-num}

%\bibliography{electrolyte}

\providecommand{\newblock}{}

\end{document}